\newlength{\widebarargwidth}
\newlength{\widebarargheight}
\newlength{\widebarargdepth}
\DeclareRobustCommand{\widebar}[1]{%
  \settowidth{\widebarargwidth}{\ensuremath{#1}}%
  \settoheight{\widebarargheight}{\ensuremath{#1}}%
  \settodepth{\widebarargdepth}{\ensuremath{#1}}%
  \addtolength{\widebarargwidth}{-0.3\widebarargheight}%
  \addtolength{\widebarargwidth}{-0.3\widebarargdepth}%
  \makebox[0pt][l]{\hspace{0.3\widebarargheight}%
    \hspace{0.3\widebarargdepth}%
    \addtolength{\widebarargheight}{0.3ex}%
    \rule[\widebarargheight]{0.95\widebarargwidth}{0.1ex}}%
  {#1}}

\newcommand{\condind}{\ensuremath{\perp\!\!\!\perp}}
\newcommand{\real}{\ensuremath{\mathbb{R}}}
\newcommand{\defn}{\ensuremath{:  =}}
\newcommand{\Thetastar}{\ensuremath{\Theta^*}}
\newcommand{\cstar}{\ensuremath{c_{\Thetastar}}}
\newcommand{\qbar}{\ensuremath{\widebar{q}}}

\newcommand{\supp}{\operatorname{supp}}
\newcommand{\Ghat}{\ensuremath{\widehat{G}}}
\newcommand{\tr}{\operatorname{tr}}
\newcommand{\Sigmahat}{\ensuremath{\widehat{\Sigma}}}
\newcommand{\E}{\ensuremath{\mathbb{E}}}
\newcommand{\Sigmastar}{\ensuremath{\Sigma^*}}
\newcommand{\VEC}{\operatorname{vec}}
\newcommand{\opnorm}[1]{\left|\!\left|\!\left|{#1}\right|\!\right|\!\right|}
\newcommand{\Sigmatil}{\ensuremath{\widetilde{\Sigma}}}
\newcommand{\cG}{\ensuremath{\mathcal{G}}}
\newcommand{\argmin}{\operatornamewithlimits{argmin}}
\newcommand{\Det}[1]{\left|{#1}\right|}

\documentclass[conference]{IEEEtran}

\usepackage[cmex10]{amsmath}
\usepackage{amssymb,amsthm,mathrsfs}
\usepackage[colorlinks]{hyperref}
\usepackage[pdftex]{graphicx}
\newtheorem{theorem*}{Theorem}   
\newtheorem{corollary*}{Corollary} 
\newtheorem{lemma*}{Lemma} 
\newtheorem*{example*}{Example}

\begin{document}
%
\title{On model misspecification and KL separation for Gaussian graphical models}

\author{\IEEEauthorblockN{Varun Jog}
\IEEEauthorblockA{Department of EECS\\
University of California at Berkeley\\
Berkeley, CA 94720\\
Email: varunjog@eecs.berkeley.edu}
\and
\IEEEauthorblockN{Po-Ling Loh}
\IEEEauthorblockA{Department of Statistics\\
The Wharton School\\
University of Pennsylvania\\
Philadelphia, PA 19104\\
Email: loh@wharton.upenn.edu}}


%


\maketitle

\begin{abstract}
We establish bounds on the KL divergence between two multivariate Gaussian distributions in terms of the Hamming distance between the edge sets of the corresponding graphical models. We show that the KL divergence is bounded below by a constant when the graphs differ by at least one edge; this is essentially the tightest possible bound, since classes of graphs exist for which the edge discrepancy increases but the KL divergence remains bounded above by a constant. As a natural corollary to our KL lower bound, we also establish a sample size requirement for correct model selection via maximum likelihood estimation. Our results rigorize the notion that it is essential to estimate the edge structure of a Gaussian graphical model accurately in order to approximate the true distribution to close precision.
\end{abstract}


%
\IEEEpeerreviewmaketitle

\section{Introduction}

Graphical models have enjoyed increasing popularity in a wide variety of scientific disciplines, including social networks~\cite{CarEtal05}, computer vision~\cite{NowLam10}, neuroscience~\cite{Hin05}, molecular biology~\cite{Fri04}, and clinical medicine~\cite{KalEtal10}. Recent years have also seen substantial theoretical advances regarding graphical models in high-dimensional statistics (see, e.g., \cite{MeiBue06, YuaLin07, FriEtal08, RavEtal11, SanWai12}). Broadly speaking, the goal of statistical estimation in graphical models is to (a) estimate the edge structure of the graph, which encodes conditional independence relationships between variables; and (b) infer the parameters of the distribution. The two goals are often treated separately: Upon determining the edges of the graph, the parameters are fit with respect to a reduced search space. This reduces the dimensionality of the subsequent parameter estimation problem, which may be advantageous in high-dimensional problems where the underlying graph is sparse. We consider a setting where data are collected in the form of joint observations; in high-dimensional scenarios, the number of nodes is assumed to be much larger than the number of observations.

However, when parameter estimation is conducted in the wake of edge estimation, inaccuracies in the estimated graph structure will propagate to the parameter estimation step. Although superfluous edges may subsequently be removed by setting the corresponding parameters to zero, missing edges in the estimated graph may lead to model misspecification. Consequently, the estimated distribution may be far from the actual distribution. Various authors (e.g., \cite{MeiBue06, DrtPer07, BreEtal08, RavEtal11, AnaEtalGaussian12}) have established sufficient conditions for specific estimation procedures that guarantee correct edge recovery, albeit under fairly stringent conditions that are more restrictive than the conditions needed for $\ell_1$- and $\ell_2$-consistency.

In this paper, we explore the following question: If the edge structure of the graph is estimated incorrectly, how large is the deviation between the true distribution and the closest fit with respect to the errant graphical model? We restrict our attention to Gaussian graphical models. Our main contribution is to establish a constant lower bound on the KL divergence between the true distribution and the closest approximation when the graphs differ by even a single edge. This should be viewed in conjunction with the work of Zhou et al.~\cite{ZhoEtal11}, who establish upper bounds on the edge discrepancy for a certain graph estimation procedure. Indeed, our result stipulates the need to identify the edge structure of the true graphical model with complete accuracy in order to approximate the underlying distribution to arbitrary precision.

Our results have interesting connections to other lines of previous work. Theorem~\ref{ThmQstar} below relates the KL divergence between two Gaussian distributions with different graphical models to the conditional mutual information between the pair of variables corresponding to the edge discrepancy. Bounds on a similar conditional mutual information expression are used to derive sample complexity results for a graphical model estimation procedure proposed by Anandkumar et al.~\cite{AnaEtalGaussian12, AnaEtalIsing12}, and indeed, our Lemma~\ref{LemMutualInfo} is similar to a proposition proved in that paper. However, rather than focusing on requirements for statistical consistency of a particular graphical model estimation algorithm, we leverage this lemma to lower-bound the KL divergence between distributions in terms of entries of the inverse covariance matrix. In a recent paper, Bresler~\cite{Bre14} provides lower bounds on the conditional mutual information between pairs of variables in an Ising model, although it is unclear whether that result could be used to derive a similar constant lower bound on the KL divergence between Ising models with differing graphical structure.

The remainder of the paper is organized as follows: In Section~\ref{SecBackground}, we provide a precise mathematical formulation of the problem under consideration and introduce relevant notation. Section~\ref{SecMain} contains statements of our main theorems, where we first lower-bound the KL divergence in terms of the conditional mutual information and then in terms of a constant parameter defined according to entries of the true inverse covariance matrix. We then discuss an easy consequence of the KL bound regarding the sample complexity of a likelihood-based approach for model selection. We close in Section~\ref{SecWarren} with some extensions of our KL lower bound and an example showing that the KL separation does not necessarily grow in a meaningful way with the Hamming distance between the edge set of the true graph and a candidate estimator. Detailed proofs may be found in the arXiv version of the manuscript~\cite{JogLoh15}.


\noindent \textbf{Notation:} For functions $f(n)$ and $g(n)$, we write $f(n) \precsim g(n)$ to indicate that \mbox{$f(n) \le c g(n)$} for some universal constant \mbox{$c \in (0, \infty)$,} and similarly, we write $f(n) \succsim g(n)$ when \mbox{$f(n) \ge c' g(n)$} for some universal constant $c' \in (0, \infty)$. We use the symbol $\condind$ to indicate independence. For a matrix $M$, we write $\opnorm{M}_F$ to denote the Frobenius norm, and let $\lambda_{\max}(M)$ denote the maximum eigenvalue of $M$. We write $M(i,j)$ to denote the $(i,j)^{\text{th}}$ entry of $M$ and $\supp(M) \defn \{(i,j): i \le j \text{ and } M(i,j) \neq 0\}$ to denote the (ordered) support of $M$. Finally, $\VEC(M)$ denotes the vectorized version of the matrix and $|M| = \det(M)$ denotes the determinant.


\section{Background and problem setup}
\label{SecBackground}

Consider a zero-mean multivariate Gaussian distribution $q_\Theta \defn N(0, \Theta^{-1})$ with inverse covariance matrix \mbox{$\Theta \in \real^{p \times p}$.} Recall that the \emph{Gaussian graphical model} corresponding to the distribution $q_\Theta$ is given by the undirected graph \mbox{$G(\Theta) = (V, E)$,} where $V = \{1, \dots, p\}$ and $E = \supp(\Theta)$ is the support of the matrix $\Theta$. This is a special case of the well-developed theory on undirected graphical models, also known as Markov random fields, where nodes represent individual variables in a joint distribution $X = (X_1, \dots, X_p)$, and missing edges represent conditional independence relationships between subsets of variables. In particular, $(i,j) \notin E$ implies that $X_i \condind X_j | X_{\{i,j\}^c}$, where we write $X_{\{i,j\}^c}$ to denote the collection of variables $\{X_1, \dots, X_p\} \setminus \{X_i, X_j\}$. For a more detailed exposition on graphical models, see Lauritzen~\cite{Lau96} or Koller and Friedman~\cite{KolFri09} and the references cited therein.

We now consider a pair of $p$-dimensional multivariate Gaussian distributions $q_1 = q_{\Theta_1}$ and $q_2 = q_{\Theta_2}$. Our main goal in this paper is to quantify the distance between $q_1$ and $q_2$ in terms of the discrepancy between $G_1 = G(\Theta_1)$ and $G_2 = G(\Theta_2)$. The distance between $q_1$ and $q_2$ is measured via the Kullback-Leibler divergence between $q_1$ and $q_2$:
\begin{equation*}
KL(q_1 || q_2) = \int_{\mathbb R^p} q_1(x) \log \frac{q_1(x)}{q_2(x)} dx.
\end{equation*}
For a fixed distribution $q_1$ defined over $G_1$, we wish to find the infimum $\inf_{q_2} KL(q_1 || q_2)$, where $q_2$ ranges over all distributions defined over $G_2$. Note that if $G_1$ is a subgraph of $G_2$, the value of this infimum may approach $0$ if we tend $\Theta_2(i,j) \to 0$ for $(i,j) \in E_2 \setminus E_1$. Hence, we insist that there is at least one edge $(i, j) \in E_1$ such that $(i,j) \notin E_2$.

Some of our results will be stated in terms of particular classes of positive definite matrices. Let\begin{multline*}
\Omega_\infty(\alpha, h) \defn \big\{\Theta \succ 0: \Theta(i, i) \le h, \quad \forall i; \text{ and} \\
|\Theta(i,j)| \ge \alpha, \quad \forall (i,j) \text{ s.t. } \Theta(i,j) \neq 0\big\},
\end{multline*}
and
\begin{equation*}
\Omega_F(\gamma) \defn \{\Theta \succ 0: \opnorm{\Theta}_F \le \gamma\},
\end{equation*}
where $\Omega_\infty$ imposes bounds on individual entries and $\Omega_F$ imposes a uniform bound on the Frobenius norm. Note that a similar class to $\Omega_\infty$, with an additional upper bound on the off-diagonal entries, was analyzed by previous authors for Ising models~\cite{SanWai12, Bre14}, and a heuristic justification for entrywise restrictions on the inverse covariance class vis-\`{a}-vis identifiability may be found in Santhanam and Wainwright~\cite{SanWai12}. As explained in the remark following Corollary~\ref{CorAlphaH} below, the ratio $\frac{\alpha}{h}$ may be viewed as a surrogate for the minimum signal strength of a multivariate Gaussian distribution with inverse covariance matrix $\Theta$. Furthermore, the fact that $\Theta$ is positive semidefinite implies that $|\Theta(i,j)| \le h$, for $i \neq j$.


\section{Main results and consequences}
\label{SecMain}

We now present our core theoretical results. We begin with the following theorem, which quantifies the KL divergence between an arbitrary distribution $q_1$ and a distribution $q_2$ taken from a class with at least one edge missing.

\begin{theorem*}
\label{ThmQstar}
Let $X = (X_1, \dots, X_p)$ be drawn from a multivariate distribution with density $q_1$. Then
\begin{equation}
\label{EqnRoti}
\min_{q_2: X_1 \condind X_2 | X_{\{1,2\}^c}} KL(q_1 || q_2) \ge I(X_1; X_2 | X_3, \dots, X_p),
\end{equation}
where $I(X_1; X_2 | X_3, \dots, X_p)$ denotes the conditional mutual information with respect to the distribution $q_1$. Equality is achieved when $q_2 = q_2^*$, where
\begin{multline*}
q_2^*(x_1, x_2, \dots, x_p) \defn q_1(x_1|x_3, \dots, x_n) \\
\cdot q_1(x_2|x_3, \dots, x_n) \cdot q_1(x_3, \dots, x_n).
\end{multline*}
\end{theorem*}

\noindent \textbf{Remark:} Note that we do not impose any distributional assumptions on either $q_1$ or $q_2$. Furthermore, if the edge $(1,2)$ is also absent in the graphical model representation of $q_1$, we have $I(X_1; X_2 | X_3, \dots, X_p) = 0$. Consequently, equality is achieved in equation~\eqref{EqnRoti} with both sides equal to 0. \\

When the variables are jointly Gaussian, it is possible to express the conditional mutual information $I(X_1; X_2 | X_3, \dots, X_p)$ cleanly in terms of the inverse covariance matrix of $q_1$. Our next result accordingly lower-bounds the KL divergence between two multivariate Gaussian distributions $q_{\Thetastar}$ and $q_\Theta$ in terms of the quantity
\begin{equation*}
\cstar \defn \min_{(i,j): \Thetastar(i,j) \neq 0} \left\{\frac{\Thetastar(i,i) \Thetastar(j,j)}{\Thetastar(i,i) \Thetastar(j,j) - \Thetastar(i,j)^2}\right\}.
\end{equation*}
Note that when $\Thetastar \succ 0$, each $2 \times 2$ submatrix of $\Thetastar$ over the indices $i$ and $j$ is also positive definite, so $\Thetastar(i,i) \Thetastar(j,j) - \Thetastar(i,j)^2 > 0$. Hence, $\cstar > 1$, and the quantity appearing in the lower bound of Theorem~\ref{ThmOneEdge} strictly positive.


\begin{theorem*}
\label{ThmOneEdge}
Consider a fixed $\Thetastar \succ 0$, and let $\Theta \succ 0$ be such that $\supp(\Thetastar) \backslash \supp(\Theta) \neq \emptyset$. Then
\begin{equation*}
KL(q_{\Thetastar} || q_\Theta) \ge \frac{1}{2} \log(\cstar).
\end{equation*}
\end{theorem*}

The proof of Theorem~\ref{ThmOneEdge}
stems from the explicit relationship between the entries of $\Thetastar$ and the conditional correlations between corresponding pairs of variables. Note that the condition $\supp(\Thetastar) \backslash \supp(\Theta) \neq \emptyset$ is necessary for the validity of the theorem; we could otherwise take $\Theta = \Thetastar$ to make the KL divergence equal to zero. 

\noindent \textbf{Remark:} From the point of view of graphical model estimation, Theorem~\ref{ThmOneEdge} provides a strong cautionary message that if \emph{at least one} edge in the true graph with edge set $\supp(\Thetastar)$ is missing, the KL divergence between $q_{\Thetastar}$ and the best possible fit is lower-bounded by the constant $\frac{1}{2} \log(\cstar) > 0$. Indeed, Theorem~\ref{ThmOneEdge} guarantees that if $G^* = G(\Thetastar)$ is the true graphical model and $G$ is any other graph with $E(G^*) \backslash E(G) \neq \emptyset$, then
\begin{equation*}
\min_{\Theta \succ 0: \; \supp(\Theta) \subseteq E(G)} KL(q_{\Thetastar} || q_\Theta) \ge \frac{1}{2} \log\left(\cstar\right).
\end{equation*}
Theorem~\ref{ThmOneEdge} is an important partner result to the theoretical conclusions of Zhou et al.~\cite{ZhoEtal11}, where an upper bound is provided on the Hamming distance between the edge sets of the true graphical model and the graphical model estimated by their algorithm. Our theorem states that whenever the Hamming distance between the edge sets is at least one, the KL divergence between the true distribution and the closest distribution in the estimated class is already bounded below by a constant. This emphasizes the importance of selecting the true edge set (or a superset thereof) when estimating the structure of the graphical model. \\

Specializing Theorem~\ref{ThmOneEdge} to the class of matrices $\Omega_\infty(\alpha, h)$, we have the following simple corollary:

\begin{corollary*}
\label{CorAlphaH}
Suppose $\Thetastar \in \Omega_\infty(\alpha, h)$. If $\Theta \succ 0$ is such that $\supp(\Thetastar) \backslash \supp(\Theta) \neq \emptyset$, then
\begin{equation*}
KL(q_{\Thetastar} || q_\Theta) \ge \frac{1}{2}\log \left(\frac{1}{1-\alpha^2/h^2}\right).
\end{equation*}
\end{corollary*}


\noindent \textbf{Remark:} Note that Corollary~\ref{CorAlphaH} only requires the \emph{true} inverse covariance matrix $\Thetastar$ to lie in $\Omega_\infty(\alpha, h)$, whereas $\Theta$ may be inside or outside the class. The conclusion of the corollary suggests that the ratio $\frac{\alpha}{h}$ may be interpreted as a type of (normalized) minimum signal strength for the true distribution $q_{\Thetastar}$. Indeed, as $\frac{\alpha}{h} \to 1$,  the KL divergence between the true distribution and all alternative distributions with the incorrect graphical structure grows unboundedly. Since the lower bound on the KL divergence increases in $\alpha$ for a fixed value of $h$, Corollary~\ref{CorAlphaH} further corroborates the notion that a type of ``strong faithfulness" condition on the true inverse covariance matrix makes the problem of edge estimation more tractable for Gaussian graphical models~\cite{UhlEtal13}. However, whereas the idea of strong faithfulness was previously introduced in order to quantify the success of specific statistical estimation algorithms, Corollary~\ref{CorAlphaH} establishes that a separation between the zero and non-zero values of $\Thetastar$ actually measures the intrinsic hardness of the graphical model selection problem in an information-theoretic sense. \\

We may observe easily from the proof of Corollary~\ref{CorAlphaH} that equality is achieved in the KL bound when a single $2 \times 2$ submatrix of $\Thetastar$ corresponding to indices $i \neq j$ has diagonal entries equal to $h$ and off-diagonals equal to $\pm \alpha$, since the parameter $\cstar$ is computed as a minimum over all $2 \times 2$ submatrices. However, as explored in more detail in Section~\ref{SecWarren}, the separation in KL divergence does \emph{not} necessarily scale with the size of the edge discrepancy between $G(\Thetastar)$ and $G(\Theta)$. In the results of that section, we provide examples where an increase in the Hamming distance between the two graphs does not substantively affect the minimum KL divergence between $q_\Thetastar$ and the best alternative model. This emphasizes the fact that $\cstar$ is not purely a local (edgewise) property, and its dependence on the conditional correlation terms appearing as entries of $\Thetastar$ takes into account the behavior of other nodes in the graph, as well. \\

Our results on KL separation also have useful consequences regarding the sample complexity of a likelihood-based model selection procedure. Suppose the true inverse covariance matrix lies in the class $\Thetastar \in \Omega_F(\gamma)$. Further suppose that we have a set of candidate graphs $\cG = \{G_0, G_1, \dots, G_M\}$, with $E(G_0) = \supp(\Thetastar)$ and $E(G_0) \backslash E(G_m) \neq \emptyset$, for all $1 \le m \le M$. In other words, $G_0$ is the graphical model of the true distribution and each of the alternative graphs is missing at least one edge.

We will analyze a maximum likelihood approach, which is equivalent to minimizing the KL divergence between the true model and another distribution in the parametric class~\cite{CovTho91}. Let
\begin{equation*}
\ell_n(\Theta) \defn - \log \det(\Theta) + \tr(\Sigmahat \Theta)
\end{equation*}
denote the negative log likelihood with respect to a distribution $q_\Theta$, where $\Sigmahat$ is the empirical covariance matrix, and let
\begin{equation*}
\ell(\Theta) \defn \E_{\Thetastar}\left[\ell_n(\Theta)\right]
\end{equation*}
denote the expected value of $\ell_n(\Theta)$ with respect to $q_{\Thetastar}$. Also define the scores
\begin{equation*}
S(G_m) \defn \min_{\stackrel{\Theta \in \Omega_F(\gamma):}{\supp(\Theta) \subseteq E(G_m)}} \left\{\ell_n(\Theta)\right\}, \qquad \forall 0 \le m \le M,
\end{equation*}
where the minimum is taken over all inverse covariance matrices with Frobenius norm bounded by $\gamma$ that are consistent with the edge structure of $G_m$. We discuss the Frobenius norm bound in the remarks following Corollary~\ref{CorSampSize}. Note that computing the score of a given graph is a tractable convex optimization program, since both the objective function and constraint set $\Omega_F(\gamma)$ are easily seen to be convex. We define the graph estimator $\Ghat =\argmin_{G_m} \{S(G_m)\}$ to be the minimum-scoring graph in the collection, where we are agnostic to the choice of graph if more than one minimizer exists. We then have the following result: 

\begin{corollary*}
\label{CorSampSize}
Suppose the data are drawn from a multivariate normal distribution with covariance matrix $\Sigmastar$, and suppose a set of candidate graphs $\cG$ is given, where \mbox{$|\supp(G_m)| \le s + p$} for all $m \ge 0$. Suppose the sample size satisfies \mbox{$n \ge \frac{4C^2\gamma^2}{\cstar^2} \cdot \lambda_{\max}^2(\Sigmastar) (p+s)\log p$.} Then with probability at least $1 - c \exp(-c' \log p)$, we have $\Ghat = G(\Thetastar)$.
\end{corollary*}

\noindent \textbf{Remark:} It is helpful to compare Corollary~\ref{CorSampSize} with the required sample size for related results on Gaussian graphical model selection. We first compare our result to the graphical model selection guarantees of Ravikumar et al.~\cite{RavEtal11}. Although the sample size scaling $n \succsim (p+s) \log p$ required by our corollary is somewhat stronger than the $n \succsim d^2 \log p$ requirement of Ravikumar et al.~\cite{RavEtal11}, where $d$ denotes the degree of the graph $G_0$, we do not impose any of the irrepresentible conditions that are rather restrictive and somewhat uninterpretable. Similarly, nodewise regression methods~\cite{MeiBue06} are consistent for model selection under the milder sample size scaling $n \succsim d \log p$, but under more stringent incoherence assumptions. Note that in our result, the constant $\cstar$ takes the role of a beta-min condition, assumed by previous authors in order to derive model selection consistency. \\

We now discuss the parameter $\gamma$ that bounds the Frobenius norm of inverse covariance matrices in our model class. This additional parameter is somewhat undesirable if  we expect the Frobenius norm to scale with $p$ (e.g., for jointly independent random variables), since it creates an even larger factor in the sample size requirement; however, it is the same assumption imposed for the purpose of Gaussian graphical model estimation in Zhou et al.~\cite{ZhoEtal11}. Some matrix norm bound on the class of inverse covariance matrices under consideration is certainly necessary, although we are unsure whether one can do better. Furthermore, it is hard to compare our Frobenius norm assumption directly with the $\ell_\infty$-operator norm bounds on population-level matrices appearing in the analyses of alternative methods~\cite{MeiBue06, RavEtal11}. We note the useful observation from Zhou et al.~\cite{ZhoEtal11} that if the diagonal entries of $\Sigmastar$ are known a priori, we may replace the estimate $\Sigmahat$ of the covariance matrix $\Sigmastar$ with the matrix $\Sigmatil$ in the definition of $\ell_n(\Theta)$, where $\Sigmatil$ has the correct diagonal entries. Then a sharper analysis leads to the slightly milder sample size requirement $n \ge \frac{4C^2 \gamma^2}{\cstar^2} \lambda_{\max}^2(\Sigmastar) s \log p$. However, the assumption that the diagonal entries are known exactly may be too strong in practical applications.


\section{Extensions and counterexamples}
\label{SecWarren}

Theorem~\ref{ThmOneEdge} shows that if the estimated graph is missing at least one edge, the KL divergence between the true and estimated distributions is bounded below by a constant. In general, we may study the problem of evaluating a lower bound $L(d)$ for the case of $d \geq 1$ missing edges. The value of $L(1)$ is given by Theorem~\ref{ThmOneEdge} and Corollary~\ref{CorAlphaH}. It is reasonable to conjecture that $L(d)$ scales with $d$; such a scaling would make it possible to relate the Hamming distance between two graphs to the KL divergence between pairs of probability distributions supported on the respective graphs. In this section, however, we show that $L(d)$ does not scale in a meaningful way with $d$. We present an explicit family of graphs for which  $L(1) \leq L(d) \leq C$, for some constant $C$ that is independent of $d$. This shows that the constant bound from Theorem~\ref{ThmOneEdge} is essentially tight.

We begin with the statement of Theorem~\ref{ThmProj}, which generalizes the result from Theorem~\ref{ThmQstar}. 

\begin{theorem*}
\label{ThmProj}
Let $X = (X_1, \dots, X_p)$ be as in Theorem~\ref{ThmQstar}. Let $\Theta_1$ be the inverse covariance matrix of $q_1$, and let $G_1 = (V, E_1)$ be the corresponding graph. Without loss of generality, consider the vertex $1$ and $d \geq 1$ of its neighbors $\{2, 3, \dots, d+1\}$. Let $G = (V, E)$, where \mbox{$E \defn \big\{(1,2), (1,3), \dots, (1,d+1)\big\}^c$.} Let $q_2$ be any distribution with the corresponding graphical model $G_2 = (V, E_2)$, such that $E_2 \subseteq E$. The following inequality holds:
\begin{equation}\label{EqnBigRoti}
KL(q_1 || q_2) \geq I(X_1; X_2, \dots, X_{d+1}| X_{d+2}, \dots, X_p).
\end{equation}
Equality is achieved when $q_2 = q_2^*$ is defined by
\begin{multline*}
q_2^*(x_1, \dots, x_n) = q(x_1| x_{d+2}, \dots, x_p) \\
\cdot q(x_2, \dots, x_{d+1}| x_{d+1}, \dots, x_p) \cdot q(x_{d+2}, \dots, x_p).
\end{multline*}
\end{theorem*}

\begin{figure}[h!]
\centering
\begin{tabular}{cc}
\includegraphics[scale = 0.35]{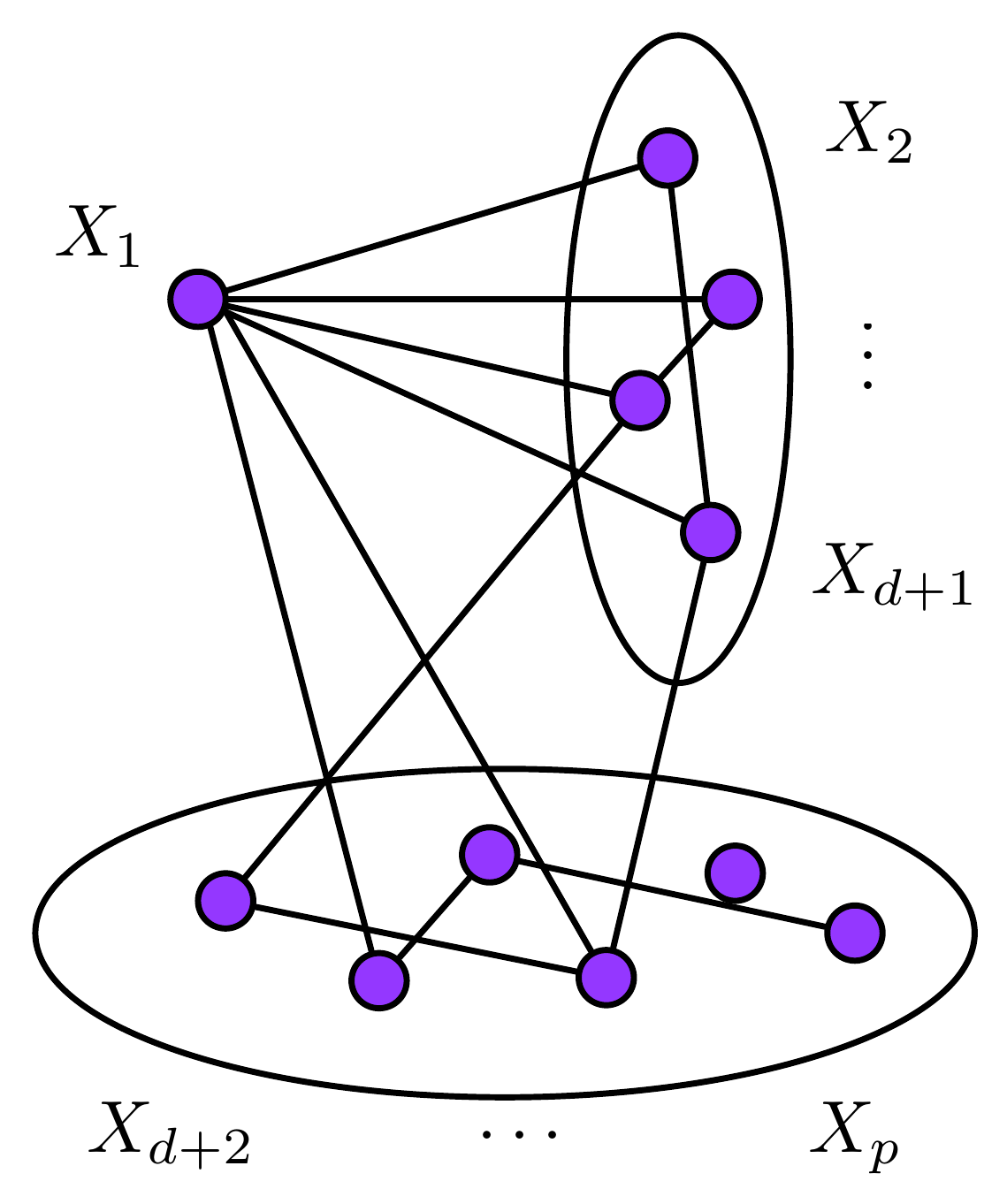} &
\includegraphics[scale = 0.35]{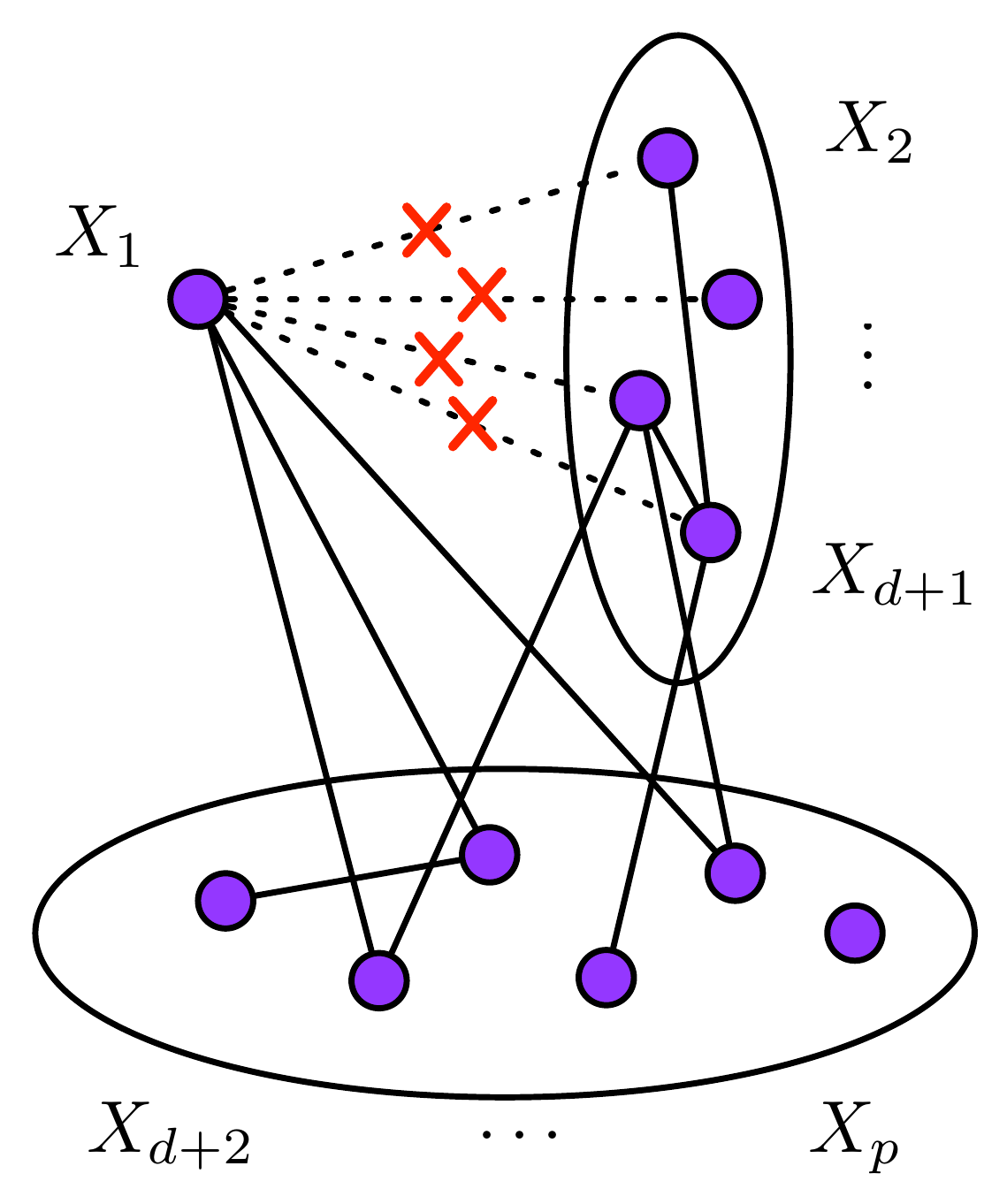} \\
(a) & (b)
\end{tabular}
\caption{An illustration of Theorem~\ref{ThmProj}. Panel (a) shows the graph $G_1 = G(\Theta_1)$, where the neighbors of node 1 include the nodes $\{2, \dots, d+1\}$. Note that node $1$ may also have other neighbors, and the remaining nodes may be connected arbitrarily. Panel (b) shows a graph $G_2$ having the property that edges $\{(1,2), \dots, (1, d+1)\}$ are missing. Again, we do not impose any restrictions on the presence or absence of other edges in the graph. Theorem~\ref{ThmProj} implies that the KL divergence between $q_1$ and any distribution $q_2$ with graphical model $G_2$ is bounded below by $I(X_1; X_2, \dots, X_{d+1}|X_{d+2}, \dots, X_p)$. 
}
\label{FigProj}
\end{figure}

\noindent An illustration of Theorem~\ref{ThmProj} is provided in Figure~\ref{FigProj}. Note that analogous to the statement of Theorem~\ref{ThmQstar}, Theorem~\ref{ThmProj} does not impose any distributional assumptions on $q_1$ or $q_2$.

\noindent \textbf{Remark:} In both Theorems~\ref{ThmQstar} and~\ref{ThmProj}, the candidate distributions $q_2$ are identified via the support of $\Theta_2$, and the particular structure of $\supp(\Theta_2)$ allows us to express $q_2$ in a convenient product form. Such a property does not hold for any arbitrary choice of $\supp(\Theta_2)$, however, although it holds for the support structures considered in Theorems~\ref{ThmQstar} and~\ref{ThmProj}. In fact, we may generalize the statement of Theorems~\ref{ThmQstar} and~\ref{ThmProj} to include any graphical structure where there exists a directed acyclic graph reflecting all conditional independence relationships present in $\supp(\Theta_2)$. \\

Using Theorem~\ref{ThmProj}, we provide an example showing that the lower bound $L(d)$ on the KL divergence for pairs of graphs differing by $d$ edges may be bounded above by a constant: 

\begin{example*}
Let $d \geq 1$. Pick a $(d+1)$-dimensional Gaussian random variable $X = (X_1, X_2, \dots, X_{d+1})$ with a distribution $q_1$, as follows: The random variables $(X_1, \dots, X_{d}, W)$ are independent standard normal random variables, and \mbox{$X_{d+1} = \sum_{i=1}^d X_i + W$.} Let the inverse covariance matrix of $q_1$ be $\Theta_1$, and let $G_1 = (V, E_1)$ denote the corresponding graph. We may check that 
\begin{equation*}
\Theta_1 = 
\begin{bmatrix}
2 &1 &\dots &1 &-1\\
1 &2 &\dots &1 &-1\\
\vdots & \vdots &\ddots &\vdots &\vdots\\ 
1 &1 &\dots &2 &-1\\
-1 &-1 &\dots &-1 &1
\end{bmatrix},
\end{equation*}
and consequently,
\begin{equation*}
\big\{ (1, 2) ,\dots, (1, d+1)\big\} \subseteq E_1.
\end{equation*}
Now choose a distribution $q_2^*$ as per Theorem \ref{ThmProj}, so
\begin{equation*}
q_2^*(x_1, \dots, x_p) = q_1(x_1)\cdot q_1(x_2, \dots, x_{d+1}).
\end{equation*}
Note that the graph $G_2$ of $q_2^*$ does not have the edges $\big\{ (1, 2) ,\dots, (1, d+1)\big\}$, so it differs from $G_1$ by at least $d$ edges. By the result of Theorem~$\ref{ThmProj}$, the KL divergence between $q_1$ and $q_2^*$ is given by
\begin{align*}
KL(q_2^* || q_1) &= I(X_1; X_2, \dots, X_{d+1})\\
&= I(X_1; X_2, \dots, X_{d}) \! + \! I(X_1; X_{d+1} | X_2, \dots, X_{d})\\
&\stackrel{(a)}= I(X_1; X_{d+1} | X_2, \dots, X_{d})\\
&\stackrel{(b)}= \frac{1}{2} \log 2,
\end{align*}
where in $(a)$, we use the fact that $X_1 \condind (X_2, \dots, X_d)$ by construction, and in $(b)$, we use Lemma~\ref{LemMutualInfo}. Our example shows that $L(d) \leq \frac{1}{2}\log 2$ for all $d \geq 1$. Note that the lower bound appearing in Theorem~\ref{ThmOneEdge} is equal to $\frac{1}{2}\log \left(\frac{4}{3}\right)$ in this case and is achieved, e.g., when only edge $(1,2)$ is removed. 
\end{example*}


\section{Discussion}

We have characterized the KL divergence between multivariate Gaussian distributions with edge discrepancies in the corresponding graphical models. Our constant-valued lower bound on the KL divergence between distributions when the graphs differ by even a single edge has both positive and negative implications: On the positive side, it provides upper bounds on the required sample complexity of model selection when presented with a collection of sparse candidate graphs containing the truth; on the negative side, our result implies that the fitted distribution will always be separated from the true distribution by a constant in terms of KL divergence when the edges are misspecified. This emphasizes the importance of selecting the correct graph when model selection and parameter estimation are performed sequentially.

Future research directions include the following: Due to the parallel results for Gaussian and Ising graphical models appearing in the literature, it would be interesting to use Theorem~\ref{ThmQstar} to derive lower bounds on the KL divergence between Ising distributions with different edge structures in terms of the parameters of the underlying distribution. We conjecture that for Ising models, the KL divergence will also be bounded below by a constant when the graphical models differ by at least one edge, although the analysis may be more complicated. Furthermore, it would be interesting to derive upper bounds  on the KL divergence between models in both the Gaussian and Ising cases, which would lead to lower bounds on the sample complexity necessary for accurate edge recovery. A smattering of such results appears in the literature, but the picture seems far from complete. On a more ambitious note, it would be interesting to rigorize the tradeoff between sample and computational complexity for parameter estimation in graphical models, since one could always use fewer samples to obtain a larger superstructure of the true edge structure, at the expense of a higher computational complexity in fitting the parameters to a larger set of estimated edges.



\section*{Acknowledgement}

The authors thank Theresa Loh and Kalpana Jog for their hospitality in Madison and Pune that fostered an efficient and productive research collaboration during the winter holiday. VJ was also supported by NSF Science \& Technology Center grant CCF-0939370, Science of Information. The authors thank the anonymous reviewers for their feedback in revising the paper.



\bibliographystyle{IEEEtran}
%

\bibliography{refs}

\newpage

\appendices


\section{Proof of Theorem~\ref{ThmQstar}}
\label{SecThmQstar}

Let $q_2$ be the density of a distribution on $X$ for which $X_1 \condind X_2 \mid X_{\{1, 2\}^c}$, and let $\qbar_1$ and $\qbar_2$ denote the marginal distributions on $(X_3, \dots, X_p)$ with respect to the distributions $q_1$ and $q_2$, respectively. We have

\begin{align*}
& KL(q_1 || q_2) = \int_{\mathbb R^p} q_1(x)\log \frac{q_1(x)}{q_2(x)}dx \\
& \quad = \int_{\mathbb R^p} q_1(x) \log \frac{q_1(x_1, x_2|x_{\{1,2\}^c})q_1(x_{\{1,2\}^c})}{q_2(x_1|x_{\{1,2\}^c})q_2(x_2|x_{\{1,2\}^c})q_2(x_{\{1,2\}^c})} dx \\
&= KL(\qbar_1 || \qbar_2) + \int_{\mathbb R^p} q_1(x) \log \frac{q_1(x_1, x_2|x_{\{1,2\}^c})}{q_2(x_1|x_{\{1,2\}^c})q_2(x_2|x_{\{1,2\}^c})} dx \\
& \qquad \qquad \ge \int_{\mathbb R^p} q_1(x) \log \frac{q_1(x_1, x_2|x_{\{1,2\}^c})}{q_2(x_1|x_{\{1,2\}^c})q_2(x_2|x_{\{1,2\}^c})} dx \\
& \qquad \qquad = \int_{\real^p} q_1(x) \log \frac{q_1(x_1, x_2 | x_{\{1,2\}^c})}{q_1(x_1 | x_{\{1,2\}^c}) q_1(x_2 | x_{\{1,2\}^c})} dx \\
& \qquad  \qquad \qquad + \int_{\real^p} q_1(x) \log \frac{q_1(x_1 | x_{\{1,2\}^c}) q_1(x_2 | x_{\{1,2\}^c})}{q_2(x_1 | x_{\{1,2\}^c}) q_2(x_2 | x_{\{1,2\}^c})} dx \\
& = I(X_1; X_2 | X_3, \dots, X_p) \\
& \qquad \qquad + \int_{\real^p} q_1(x) \log \frac{q_1(x_1 | x_{\{1,2\}^c}) q_1(x_2 | x_{\{1,2\}^c})}{q_2(x_1 | x_{\{1,2\}^c}) q_2(x_2 | x_{\{1,2\}^c})} dx.
\end{align*}
Note that the conditional mutual information is constant with respect to $q_1$. We claim that the second term is always nonnegative. Indeed, we may break up the term as
\begin{align*}
& \int_{\real^p} q_1(x) \log \frac{q_1(x_1 | x_{\{1,2\}^c})}{q_2(x_1 | x_{\{1,2\}^c})} dx \\
& \qquad \qquad + \int_{\real^p} q_1(x) \log \frac{q_1(x_2 | x_{\{1,2\}^c})}{q_2(x_2 | x_{\{1,2\}^c})} dx \\
& \qquad = \int_{\real^{p-1}} q_1(x_{\{2\}^c}) \log \frac{q_1(x_1 | x_{\{1,2\}^c})}{q_2(x_1 | x_{\{1,2\}^c})} dx_1 dx_3 \cdots dx_p \\
& \qquad \quad + \int_{\real^{p-1}} q_1(x_{\{1\}^c}) \log \frac{q_1(x_2 | x_{\{1,2\}^c})}{q_2(x_2 | x_{\{1,2\}^c})} dx_2 dx_3 \cdots dx_p \\
& = \int \left(\int_\real q_1(x_1 | x_{\{1,2\}^c}) \log \frac{q_1(x_1 | x_{\{1,2\}^c})}{q_2(x_1 | x_{\{1,2\}^c})} dx_1\right) dq_1(x_{\{1,2\}^c}) \\
& + \int \left(\int_\real q_1(x_2 | x_{\{1,2\}^c}) \log \frac{q_1(x_2 | x_{\{1,2\}^c})}{q_2(x_2 | x_{\{1,2\}^c})} dx_2\right) dq_1(x_{\{1,2\}^c}). \\
\end{align*}
Finally, note that the two inner integrals are expressions for the KL divergence between the conditional distributions of $X_1 | X_{\{1,2\}^c}$ and $X_2 | X_{\{1,2\}^c}$, when $X$ is distributed according to $q_1$  and $q_2$, respectively. Hence, both integrals are nonnegative. We conclude that inequality~\eqref{EqnRoti} holds.

In order for equality to be satisfied, note that we require $KL(\qbar_1 || \qbar_2) = 0$ and the conditional KL terms to be equal to 0 for each value of $(x_3, \dots, x_p)$, meaning
\begin{equation*}
q_2(x_1 | x_3, \dots, x_p) \equiv q_1(x_1 | x_3, \dots, x_p),
\end{equation*}
and
\begin{equation*}
q_2(x_2 | x_3, \dots, x_p) \equiv q_1(x_1 | x_3, \dots, x_p).
\end{equation*}
This uniquely defines the distribution $q_2^*$.


\section{Proof of Theorem~\ref{ThmOneEdge}}
\label{SecThmOneEdge}

We begin by proving the following lemma, which we derive via a direct computation. A similar result may be found in Proposition 17 of Anandkumar et al.~\cite{AnaEtalIsing12}, but we provide the full details here for completeness. 

\begin{lemma*}\label{LemMutualInfo}
Let $X = (X_1, \dots, X_p)$ be drawn from a multivariate normal distribution with inverse covariance matrix $\Theta$. Then
\begin{multline*}
I(X_1; X_2 | X_3, \dots, X_p) \! = \! \frac{1}{2} \log \left(\frac{\Theta(1,1)\Theta(2,2)}{\Theta(1,1)\Theta(2,2)-\Theta(1,2)^2}\right),
\end{multline*}
where the mutual information is computed with respect to $q_\Theta$.
\end{lemma*}

\begin{IEEEproof}
We begin with some notation. Let $(X_1, X_2) = U$ and $(X_3, \dots, X_p) = V$. Let the covariances of $X$, $U$, and $V$ be denoted by $\Sigma_{XX}$, $\Sigma_{UU}$, and $\Sigma_{VV}$ respectively. The cross-covariance of $U$ and $V$ is denoted by $\Sigma_{UV}$. Note that 
\begin{equation*}
\Sigma_{UV} = 
\begin{bmatrix}
\Sigma_{X_1V}\\
\Sigma_{X_2V},
\end{bmatrix}
\end{equation*}
where $\Sigma_{X_iV}$ stand for the cross covariance matrices of $X_i$ and $V$, for $i \in \{1,2\}$. We have
\begin{align*}
\Sigma_{XX} 
&= 
\begin{bmatrix}
\Sigma_{UU} &\Sigma_{UV}\\
\Sigma_{UV}^T &\Sigma_{VV}
\end{bmatrix} = 
\begin{bmatrix}
\Sigma_{UU}(1,1) &\Sigma_{UU}(1,2) &\Sigma_{X_1V}\\
\Sigma_{UU}(2,1) &\Sigma_{UU}(2,2) &\Sigma_{X_2V}\\
\Sigma_{X_1V}^T &\Sigma_{X_2V}^T &\Sigma_{VV}
\end{bmatrix}.
\end{align*}

Since $(X_1, \dots, X_p)$ are jointly Gaussian, the mutual information term may be computed as 
\begin{align*}
& I(X_1; X_2 | X_{\{1,2\}^c}) = H(X_1 | X_{\{1,2\}^c}) + H(X_2 | X_{\{1,2\}^c}) \\
& \qquad \qquad \qquad \qquad \qquad \qquad - H(X_1, X_2 | X_{\{1,2\}^c})\\
& \qquad \qquad \qquad \qquad = H(X_1 | V) + H(X_2 | V) - H(U | V)\\
& \qquad \qquad \qquad \qquad = \frac{1}{2} \log \Det{\Sigma_{UU}(1,1) - \Sigma_{X_1V}\Sigma_{VV}^{-1}\Sigma_{X_1V}^T}\\
& \qquad \qquad \qquad \qquad + \frac{1}{2} \log \Det{\Sigma_{UU}(2,2) - \Sigma_{X_2V}\Sigma_{VV}^{-1}\Sigma_{X_2V}^T}\\
& \qquad \qquad \qquad \qquad -\frac{1}{2} \log \Det{\Sigma_{UU} - \Sigma_{UV} \Sigma_{VV}^{-1} \Sigma_{UV}^T}.
\end{align*}
Given a block matrix 
\begin{equation*}
M = 
\begin{bmatrix}
A &B\\
C &D
\end{bmatrix},
\end{equation*}
the Schur complement of $D$ is given by $A - BD^{-1}C$, and $\Det{A - BD^{-1}C} = \frac{\Det M}{\Det D}$.

Note that $\Sigma_{UU}(1,1) - \Sigma_{X_1V}\Sigma_{VV}^{-1}\Sigma_{X_1V}^T$ is the Schur complement of $\Sigma_{VV}$ in the matrix $\Sigma_{XX}^{(2)}$, which is $\Sigma_{XX}$ with the second row and second column removed. Similarly, $\Sigma_{UU}(2,2) - \Sigma_{X_2V}\Sigma_{VV}^{-1}\Sigma_{X_2V}^T$ is the Schur complement of $\Sigma_{VV}$ in $\Sigma_{XX}^{(1)}$, which is $\Sigma_{XX}$ with the first row and first column removed. The final term $\Sigma_{UU} - \Sigma_{UV} \Sigma_{VV}^{-1} \Sigma_{UV}^T$ is the Schur complement of $\Sigma_{VV}$ in $\Sigma_{XX}$; by the block matrix inversion formula, it equals 
\begin{equation*}
\begin{bmatrix}
\Theta(1,1) &\Theta(1,2)\\
\Theta(2,1) &\Theta(2,2)\\
\end{bmatrix}.
\end{equation*}
Thus, we obtain
\begin{align*}
I(X_1; X_2 | X_{\{1,2\}^c})  &= \frac{1}{2} \log \frac{\Det{\Sigma_{XX}^{(2)}}}{\Det{\Sigma_{VV}}} + \frac{1}{2} \log \frac{\Det {\Sigma_{XX}^{(1)}}}{\Det {\Sigma_{VV}}} \\
& \qquad -\frac{1}{2} \log \Det{\begin{bmatrix}
\Theta(1,1) &\Theta(1,2)\\
\Theta(2,1) &\Theta(2,2)\\
\end{bmatrix}}\\
&= \frac{1}{2} \log \left(\frac{\Theta(1,1)\Theta(2,2)}{\Theta(1,1)\Theta(2,2)-\Theta(1,2)^2}\right).
\end{align*}
\end{IEEEproof}

Now we are ready to derive the main result. Note that by assumption, there exists $(i,j) \in \supp(\Thetastar) \backslash \supp(\Theta)$ with $i \neq j$. Hence, $X_i \condind_\Theta X_j \mid X_{\{i,j\}^c}$, where $\condind_\Theta$ denotes conditioning with respect to the distribution $q_\Theta$. Then
\begin{equation*}
KL(q_{\Thetastar} || q_\Theta) \ge I(X_i; X_j | X_{\{i, j\}^c}) \ge \frac{1}{2} \log(\cstar),
\end{equation*}
where the first inequality follows by Theorem~\ref{ThmQstar} and the second inequality by Lemma~\ref{LemMutualInfo}, and the mutual information is computed with respect to $q_{\Thetastar}$. This is the desired inequality.


\section{Proof of Corollary~\ref{CorAlphaH}}
\label{SecCorAlphaH}

Note that since $\Theta \succ 0$ for all $\Theta \in \Omega_\infty(\alpha, h)$, we have $\alpha < h$. Hence,
\begin{align*}
& \min_{\Thetastar \in \Omega_\infty(\alpha, h)} \cstar \\
& \quad = \min_{\Thetastar \in \Omega_\infty(\alpha, h)} \left\{\min_{\Thetastar \in \Omega_\infty(\alpha, h)} \left(1 - \frac{\Thetastar(i,j)^2}{\Thetastar(i,i) \Thetastar(j,j)}\right)^{-1} \right\} \\
& \quad \ge \frac{1}{1 - \alpha^2/h^2}.
\end{align*}
The result is then an immediate consequence of Theorem~\ref{ThmOneEdge}.


\section{Proof of Corollary~\ref{CorSampSize}}
\label{SecCorSampSize}

Note that
\begin{equation*}
S(G_0) \le \ell_n(\Thetastar) \le \ell(\Thetastar) + |\ell_n(\Thetastar) - \ell(\Thetastar)|,
\end{equation*}
and for $m \ge 1$, we have
\begin{multline*}
S(G_m) \ge \min_{\stackrel{\Theta \in \Omega_F(\gamma):}{\supp(\Theta) \subseteq E(G_m)}} \{\ell(\Theta)\}  \\
- \max_{\stackrel{\Theta \in \Omega_F(\gamma):}{\supp(\Theta) \subseteq E(G_m)}} |\ell_n(\Theta) - \ell(\Theta)|.
\end{multline*}
Also note that
\begin{align*}
|\ell_n(\Theta) - \ell(\Theta)| &= \left|\tr(\Sigmahat \Theta) - \tr(\Sigmastar \Theta)\right| \\
& \le \|\VEC(\Sigmahat - \Sigmastar)\|_\infty \|\VEC(\Theta)\|_1,
\end{align*}
where $\Sigmastar \defn \Theta^{*-1}$. Furthermore, for all $\Theta$ satisfying $\supp(\Theta) \subseteq E(G_m)$ for some $m \ge 0$, we have
\begin{align*}
\|\VEC(\Theta)\|_1 & \le \sqrt{p+s} \; \|\VEC(\Theta)\|_2 \\
& = \sqrt{p+s} \; \|\Theta\|_F \le \gamma \sqrt{p+s},
\end{align*}
by the Cauchy-Schwarz inequality. Also,
\begin{equation*}
\|\VEC(\Sigmahat - \Sigmastar)\|_\infty \le C \lambda_{\max}(\Sigmastar) \sqrt{\frac{\log p}{n}},
\end{equation*}
with probability at least $1 - c \exp(-c' \log p)$, using standard Gaussian tail bounds~\cite{Ver12}. It follows that
\begin{equation*}
S(G_0) \le \ell(\Thetastar) + C \gamma \lambda_{\max}(\Sigmastar) \sqrt{\frac{(p+s)\log p}{n}},
\end{equation*}
and
\begin{equation*}
S(G_m) \ge \!\!\! \min_{\stackrel{\Theta \in \Omega_F(\gamma):}{\supp(\Theta) \subseteq E(G_m)}} \!\!\!\!\!\!\! \{\ell(\Theta)\} \, - \, C\gamma \lambda_{\max}(\Sigmastar) \sqrt{\frac{(p+s)\log p}{n}},\end{equation*}
for all $m \ge 1$, with the same probability. Finally, by Theorem~\ref{ThmOneEdge}, we have
\begin{equation*}
\min_{\stackrel{\Theta \in \Omega_F(\gamma):}{\supp(\Theta) \subseteq E(G_m)}} \{\ell(\Theta)\} - \ell(\Thetastar) \ge \cstar, \qquad \forall 1 \le m \le M.
\end{equation*}
We conclude that if $2C\gamma \lambda_{\max}(\Sigmastar) \sqrt{\frac{(p+s)\log p}{n}} < \cstar$, then $\Ghat = G_0$. Rearranging yields the desired sample size requirement.


\section{Proof of Theorem~\ref{ThmProj}}
\label{SecThmProj}

Let $\qbar_1$ and $\qbar_2$ denote the marginal distributions on $(x_{d+2}, \dots, x_p)$ with respect to the distributions $q_1$ and $q_2$, respectively. For $i < j$, let $x_{i:j}$ denote the $(j-i+1)$-dimensional vector $(x_i, \dots, x_j)$. Then

\begin{align*}
& KL(q_1 || q_2) = \int_{\mathbb R^p} q_1(x)\log \frac{q_1(x)}{q_2(x)}dx \\
&= \int q_1(x) \log \frac{q_1(x_{1:(d+1)} |x_{(d+2):p})q_1(x_{(d+2):p})}{q_2(x_1|x_{(d+2):p})q_2(x_{2:(d+1)} | x_{(d+2):p})q_2(x_{(d+2):p})} dx \\
& = KL(\qbar_1 || \qbar_2) \\
& \qquad + \int_{\mathbb R^p} q_1(x) \log \frac{q_1(x_{1:(d+1)} |x_{(d+2):p})}{q_2(x_1|x_{(d+2):p}) q_2(x_{2:(d+1)} | x_{(d+2):p})} dx \\
& \ge \int_{\mathbb R^p} q_1(x) \log \frac{q_1(x_{1:(d+1)} |x_{(d+2):p})}{q_2(x_1|x_{(d+2):p}) q_2(x_{2:(d+1)} | x_{(d+2):p})} dx \\
& = \int_{\mathbb R^p} q_1(x) \log \frac{q_1(x_{1:(d+1)} |x_{(d+2):p})}{q_1(x_1|x_{(d+2):p}) q_1(x_{2:(d+1)} | x_{(d+2):p})} dx \\
& \qquad  + \int_{\real^p} q_1(x) \log \frac{q_1(x_1|x_{(d+2):p})q_1(x_{2:(d+1)} | x_{(d+2):p})}{q_2(x_1|x_{(d+2):p}) q_2(x_{2:(d+1)} | x_{(d+2):p})} dx \\
& = I(X_1; X_{2:(d+1)} | X_{(d+2):p})\\
& \qquad + \int_{\real^p} q_1(x) \log \frac{q_1(x_1|x_{(d+2):p}) q_1(x_{2:(d+1)} | x_{(d+2):p})}{q_2(x_1|x_{(d+2):p}) q_2(x_{2:(d+1)} | x_{(d+2):p})} dx.
\end{align*}
Note that the conditional mutual information is constant with respect to $q_1$. We claim that the second term is always nonnegative. Indeed, we may break up the term as
\begin{align*}
& \int_{\real^p} q_1(x) \log \frac{q_1(x_1|x_{(d+2):p})}{q_2(x_1|x_{(d+2):p})} dx \\
& \qquad \qquad + \int_{\real^p} q_1(x) \log \frac{q_1(x_{2:(d+1)} | x_{(d+2):p})} {q_2(x_{2:(d+1)} | x_{(d+2):p})} dx \\
& \qquad = \int_{\real^{p-d}} q_1(x_1, x_{(d+2):p}) \log \frac{q_1(x_1|x_{(d+2):p})}{q_2(x_1|x_{(d+2):p})} dx_1 dx_{(d+2):p} \\
& \qquad \qquad + \int_{\real^{p-1}} q_1(x_{2:p}) \log \frac{q_1(x_{2:(d+1)} | x_{(d+2):p})}{q_2(x_{2:(d+1)} | x_{(d+2):p})} dx_{2:p} \\
& = \int \left(\int q_1(x_1 | x_{(d+2):p} \log \frac{q_1(x_1 | x_{(d+2):p})}{q_2(x_1 | x_{(d+2):p})} dx_1\right) dq_1(x_{(d+2):p}) \\
&  + \int \Big(\int q_1(x_{2:(d+1)} | x_{(d+2):p}) \\
& \qquad \qquad \cdot \log \frac{q_1(x_{2:(d+1)} | x_{(d+2):p})} {q_2(x_{2:(d+1)} | x_{(d+1):p})} dx_{2:(d+1)}\Big) dq_1(x_{(d+2):p}). \\
\end{align*}
Finally, note that the two inner integrals are expressions for the KL divergence between the conditional distributions
\begin{equation*}
X_1 \mid (X_{(d+2):p} = x_{(d+2):p})
\end{equation*}
and
\begin{equation*}
X_{2:(d+1)} \mid (X_{(d+2):p} = x_{(d+2):p}),
\end{equation*}
when $X$ is distributed according to $q_1$  and $q_2$, respectively. Hence, both integrals are nonnegative. We conclude that inequality \eqref{EqnBigRoti} holds.

Note that equality is achieved exactly when \mbox{$KL(\qbar_1 || \qbar_2) = 0$} and the conditional KL terms are equal to 0 for each value of $(x_{d+2}, \dots, x_p)$. Then
\begin{equation*}
q_2(x_1 | x_{d+2}, \dots, x_p) \equiv q_1(x_1 | x_{d+2}, \dots, x_p), 
\end{equation*}
and
\begin{multline*}
q_2(x_2, \dots, x_{d+1} | x_{d+2}, \dots, x_p) \\
\equiv q_1(x_2, \dots, x_{d+1} | x_{d+2}, \dots, x_p),
\end{multline*}
which uniquely determines the distribution $q_2^*$.

\end{document}